\documentclass[aps,prl,twocolumn,superscriptaddress,notitlepage,showpacs]{revtex4}

\usepackage{amssymb,amsfonts,amsmath} 
\usepackage[pdftex]{graphicx} 
\usepackage{epstopdf}
\usepackage{pdfpages}

\usepackage[utf8]{inputenc}
\usepackage[T1]{fontenc}
\usepackage{color}

\begin{document}


\title{Spatial Coherence Properties of Organic Molecules Coupled to Plasmonic Surface Lattice Resonances in the Weak and Strong Coupling Regimes}



\author{L.\ Shi}
\affiliation{COMP Centre of Excellence, Department of Applied Physics, Aalto University, FI-00076 Aalto, Finland}

\author{~T.~K.\ Hakala} 
\affiliation{COMP Centre of Excellence, Department of Applied Physics, Aalto University, FI-00076 Aalto, Finland} 

\author{~H.~T.\ Rekola} 
\affiliation{COMP Centre of Excellence, Department of Applied Physics, Aalto University, FI-00076 Aalto, Finland} 

\author{~J.-P.\ Martikainen} 
\affiliation{COMP Centre of Excellence, Department of Applied Physics, Aalto University, FI-00076 Aalto, Finland}

\author{~R.~J.\ Moerland}
\affiliation{COMP Centre of Excellence, Department of Applied Physics, Aalto University, FI-00076 Aalto, Finland} 
\affiliation{Department of Imaging Physics, Faculty of Applied Sciences,
Delft University of Technology, Lorentzweg 1, NL-2628 CJ, Delft, The Netherlands}

\author{P.\ T\"orm\"a}
\altaffiliation{Electronic address: paivi.torma@aalto.fi}
\affiliation{COMP Centre of Excellence, Department of Applied Physics, Aalto University, FI-00076 Aalto, Finland} 




\begin{abstract}
We study spatial coherence properties of a system composed of periodic silver nanoparticle arrays covered with a fluorescent organic molecule (DiD) film. The evolution of spatial coherence of this composite structure from the weak to the strong coupling regime is investigated by systematically varying the coupling strength between the localized DiD excitons and the collective, delocalized modes of the nanoparticle array known as surface lattice resonances. A gradual evolution of coherence from the weak to the strong coupling regime is observed, with the strong coupling features clearly visible in interference fringes. A high degree of spatial coherence is demonstrated in the strong coupling regime, even when the mode is very excitonlike (80 $\%$), in contrast to the purely localized nature of molecular excitons. We show that coherence appears in proportion to the weight of the plasmonic component of the mode throughout the weak-to-strong coupling crossover, providing evidence for the hybrid nature of the normal modes. 
\end{abstract}


\maketitle



Spatial coherence properties of waves can be probed by passing a wave front through distant slits and observing interference. Inspired by this phenomenon well known for classical radiation, interference experiments were crucial in establishing the wave-particle nature of single photons, as well as massive particles \cite{Davisson, Arndt, Juffmann}, within quantum mechanics. In the experiments \cite{Davisson, Arndt, Juffmann} the quantum mechanical wave properties of matter became visible at low temperatures. Here, we consider a different question: the spatial coherence properties of objects, or modes, that are hybrids of wavelike and particlelike components. Mixing a localized matter component with light may possibly give the hybrid object a nontrivial spatial coherence length. 

Examples of light-matter hybrids include coherent superpositions of atoms and cavity photons \cite{Rempe1987,Thompson1992}, semiconductor cavity polaritons, which have been brought to quantum degeneracy and condensation \cite{Kasprzak}, and cavity photon mediated strong coupling between spatially separated localized molecular excitons \cite{Lidzey}. Recently, delocalized electromagnetic modes supported by metal surfaces (surface plasmon polaritons) or periodic arrays of metallic nanoparticles [surface lattice resonances (SLRs) \cite{Schatz, Abajo2007, Auguie, Zhou}] have been shown to strongly couple with localized emitters \cite{Bellessa, Dintinger2005, Hakala, Gomez, Schwartz2011,Vasa2013,Rodriguez2,Aaro}. The strong coupling in these plasmonic systems involves a large number $N$ of emitters. The normal mode splittings observed are consistent both with classical linear dispersion theory and with the vacuum Rabi splitting obtained as the low excitation limit of the Dicke model, similarly to the early experiments on many atoms in cavities \cite{Zhu1990}. The collective behavior of many emitters has been clearly demonstrated in these systems, manifested as the $\sqrt{N}$ dependence of the splitting. The observed splittings in dispersions strongly support the interpretation that the new normal modes are hybrid modes formed by strong coupling of lightlike (the surface plasmon polariton/SLR) and matterlike (the molecular excitation) components. Observations of the dispersions alone, however, cannot directly test whether the new modes carry all the essential properties of the original modes, as should be the case if the hybrids are linear, coherent combinations of the original modes. In particular, spatial coherence is the specific characteristic of an extended light mode: in order to prove that the new modes carry this property, interference experiments are needed. To be conclusive, it is necessary to show that the coherence appears in proportion to the weight of the light mode in the hybrid. This in turn requires a systematic study of coherence throughout the weak-to-strong coupling crossover. This is the goal of the present work.

The spatial interference effects of light-matter hybrids have been studied in a few experiments in the context of exciton-polariton condensates \cite{Richard, Kasprzak, Deng}. In plasmonic systems, only one experiment has been reported \cite{Bellessa2}: signatures of coherence were observed in the strong coupling regime in a planar metal surface---molecular film system. However, that work does not prove the connection of the spatial coherence with the weight of the light component since there was no study of the weak-to-strong coupling crossover (a different system, namely quantum dots, was given as the weak coupling reference). Here, we study the spatial coherence properties of a system composed of periodic silver nanoparticle arrays covered with fluorescent organic molecules (DiD) by employing a double slit experiment. We gradually increase the molecule concentration to investigate both the strong and the weak coupling coherence properties within the same system. 

\begin{figure}[ht!]
\centerline{\includegraphics[width=1\columnwidth]{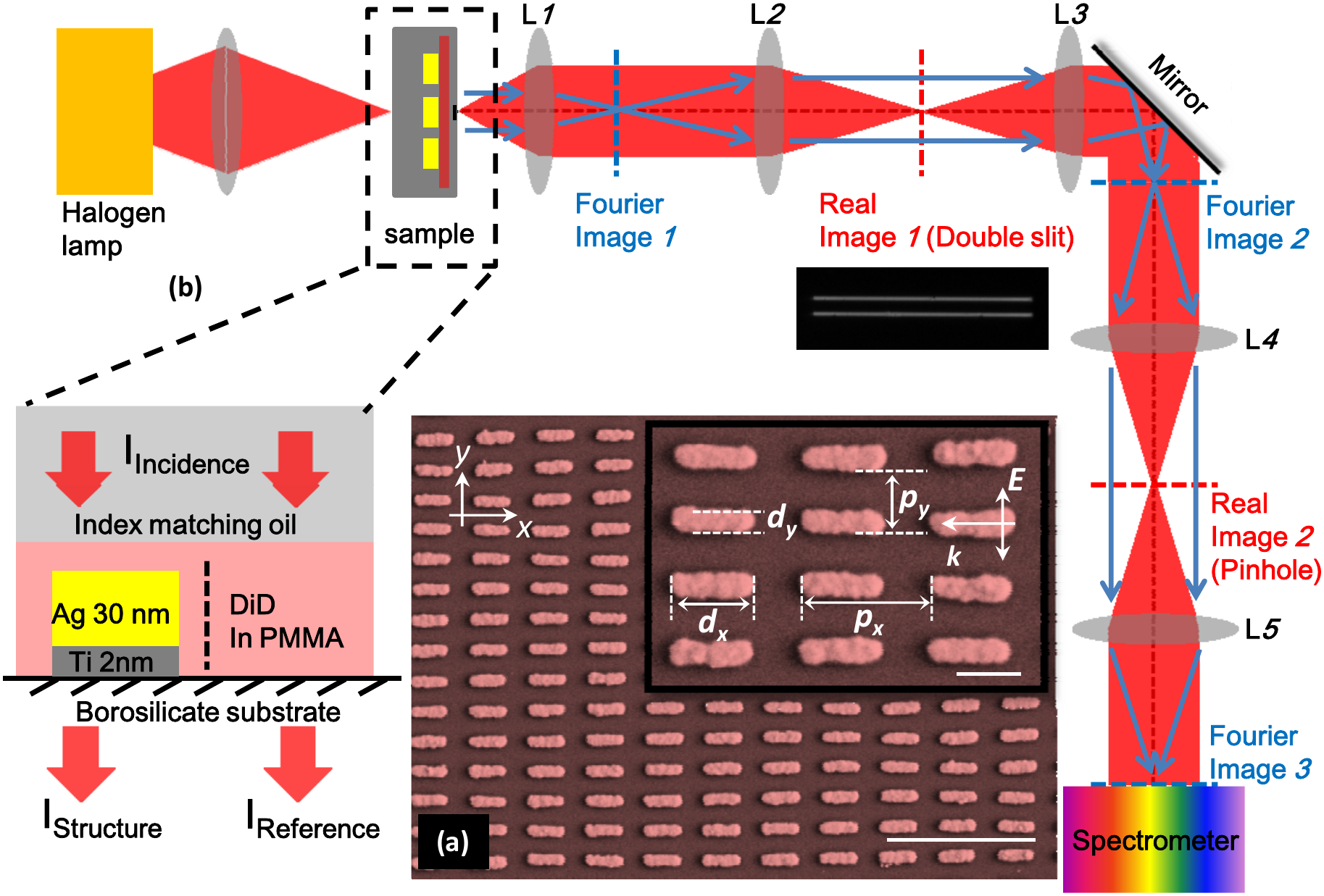}}
\caption{(a) A SEM of a typical sample. The scale bar is 1 $\mu$m (200 nm for the inset).
(b) The measurement setup. Angle resolved transmission spectra for each array were measured by placing the back focal plane of the sample at the entrance slit of the spectrometer. For spatial coherence measurements, a double slit was placed at the first intermediate image plane of the system.
}
\label{fig:Setup}
\end{figure}

Figure~\ref{fig:Setup}(a) shows a scanning electron micrograph (SEM) of a typical array (for fabrication details, see Supplemental Material \cite{Supplementarynote}). The $d_y$ = 50 nm, $p_y$ = 200 nm, while $d_x$ and $p_x$ were varied between 133--400 nm and 380--500 nm, respectively. The DiD concentration in poly(methyl methacrylate) film was varied between 20--800 mM. 

The measurement setup is depicted in Fig.~\ref{fig:Setup}(b). $y$-polarized white light was incident on the sample; see Fig.~\ref{fig:Setup}(a). Angle and wavelength-resolved transmission spectra $T = I_{\textrm{Structure}} / I_{\textrm{Reference}}$ [Fig.~\ref{fig:Setup}(b)] were measured and subsequently used for calculating the dispersion for each array. The entrance slit of the spectrometer and the in-plane wave vector $k$ is parallel to the $x$ axis of the sample with magnitude $k = 2 \pi / \lambda \sin (\theta)$, where $\lambda$ is the wavelength in the medium and $\theta$ is the angle between the optical axis and the light propagation direction.

\begin{figure*}[ht!]
\includegraphics[width=1.0\textwidth]{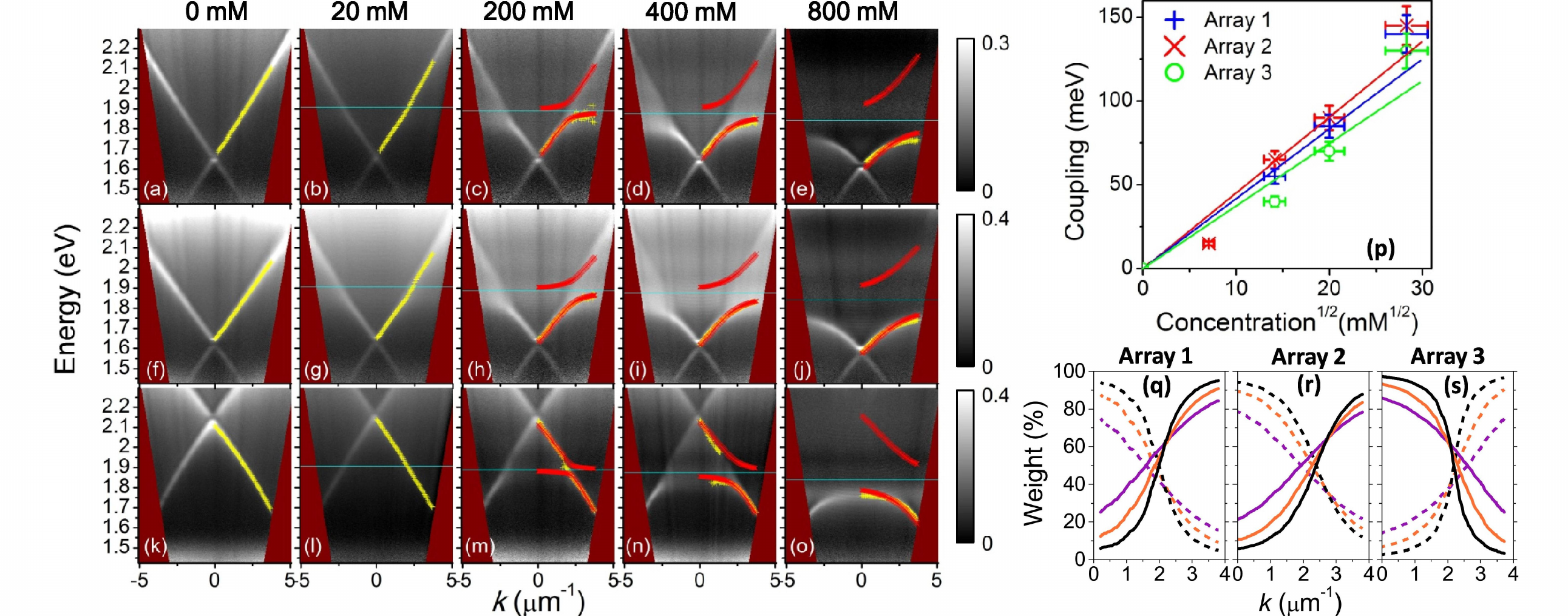}
\caption{The dispersions of three different nanoparticle arrays with inreasing DiD concentration. (a)--(e) Array 1, $(d_x) \times (d_y)$ = 50 nm $\times$ 220 nm, $p_x$ = 500 nm. (f)--(j) Array 2, $(d_x) \times (d_y)$ = 50 nm $\times$ 355 nm, $p_x$ = 500 nm. (k)--(o) Array 3, $(d_x) \times (d_y)$ = 50 nm $\times$ 167 nm, $p_x$ = 380 nm. The first column corresponds to a case without DiD molecules, while the second, third, fourth, and fifth columns have 20, 200, 400, and 800 mM concentrations of DiD, respectively. White areas correspond to maximum extinction. The blue horizontal lines depict the absorption maximum of the DiD film. The yellow lines correspond to peak positions obtained from fitting a Gaussian curve to the line cuts of dispersions while keeping $k$ constant, and the red lines are obtained from the coupled oscillator model. (p) The SLR-exciton coupling strength as a function of square root of concentration. The blue plus signs, red crosses, and green circles correspond to arrays 1, 2 and 3, respectively. (q)-(s) The relative SLR-exciton weights of the arrays 1-3, respectively. The solid (dashed) line corresponds to exciton (SLR) percentage and black, orange and purple to concentrations of 200, 400, and 800 mM, respectively.
}
\label{fig:Dispersion}
\end{figure*}

In Figs.~\ref{fig:Dispersion}(a)--\ref{fig:Dispersion}(o) are shown the measured angle resolved extinction ($1 - T$) spectra for different nanoparticle arrays. Several observations can be made from these figures. First, the energy of the $\Gamma$ point ($k = 0$) can be changed by changing the periodicity [see for example Figs.~\ref{fig:Dispersion}(a) and \ref{fig:Dispersion}(k)]. Second, upon coupling of the <+1, 0> and <-1, 0> diffractive orders \cite{Barnes}, a band gap is formed in Fig.~\ref{fig:Dispersion}(f) and the associated new modes can be made either dark or bright by changing the filling fraction [$d_x/p_x$, see Fig.~\ref{fig:Setup}(a)]. For details, see Supplemental Material \cite{Supplementarynote}. 

The dispersions in Fig.~\ref{fig:Dispersion}(b)--\ref{fig:Dispersion}(e) illustrate how the system gradually evolves from the weak to the strong coupling regime with increasing molecular concentration. A clear modification of the system energies is observed in Figs.~\ref{fig:Dispersion}(b)--\ref{fig:Dispersion}(d), which then in Fig.~\ref{fig:Dispersion}(e) develops into a distinctive band bending and anticrossing at the energy corresponding to the absorption maximum of the molecule, a behavior that is characteristic for the strong coupling regime. Similar evolution from weak to strong coupling regime can readily be identified for arrays 2 [Figs.~\ref{fig:Dispersion}(f)--\ref{fig:Dispersion}(j)] and 3 [Figs.~\ref{fig:Dispersion}(k)--\ref{fig:Dispersion}(o)], but now the system energies are drastically different due to different filling fraction (array 2) and periodicity (array 3). These results demonstrate how the choice of geometry and molecular concentration provides excellent control over the system properties. 

In the strong coupling theory, the new modes are linear combinations of the uncoupled SLRs and the molecular excitations. To describe such hybrid modes, we employ a coupled oscillator model satisfying the equation
\begin{equation}
\left(
\begin{array}{cc}
E_{SLR} (k)+ i \gamma_{SLR} & \Omega \\
\Omega & E_{DiD} + i \gamma_{DiD}
\end{array} \right)
\left(
\begin{array}{c}
\alpha \\
\beta
\end{array} \right)
= 0, \label{eq:matrix}
\end{equation}
where $E$ and $\gamma$ are the energies and the widths of the uncoupled modes, $\Omega$ is the coupling strength between the SLR and DiD, and $\alpha$ and $\beta$ are the coefficients of the linear combination of SLR and the DiD exciton (for details see Supplemental Material \cite{Supplementarynote}). The SLR-exciton coupling strength $\Omega$ and the linewidth $\gamma_{\mathrm{DiD}}$ of the exciton are used as fitting parameters. The resulting mode energies are plotted in Figs.~\ref{fig:Dispersion}(c)--\ref{fig:Dispersion}(e), \ref{fig:Dispersion}(h)--\ref{fig:Dispersion}(j), and \ref{fig:Dispersion}(m)--\ref{fig:Dispersion}(o) for different arrays and are in good agreement with the experimentally observed mode energies. The SLR-exciton coupling is expected to scale as $\sqrt{N/V}$, where $N$ is the number of molecules and $V$ is the mode volume \cite{Agranovich, Garcia}: this is confirmed in Fig.~\ref{fig:Dispersion}(p). Notably, the size of the observed splitting is in reasonable agreement with microscopic theory \cite{Agranovich} (see Supplemental Material \cite{Supplementarynote}). Note that spectrally broad emitters coupled to spectrally selective (plasmon) modes can produce luminescence spectra reminiscent of those observed in strongly coupled systems (see, e.g., Ref. \cite{Krenn}). That we observe strong coupling instead of this phenomenon is proven by the series of different concentrations that we studied, showing the $\sqrt{N/V}$ dependence expected for strong coupling.

\begin{figure*}[ht!]
\centerline{\includegraphics[width=2.0\columnwidth]{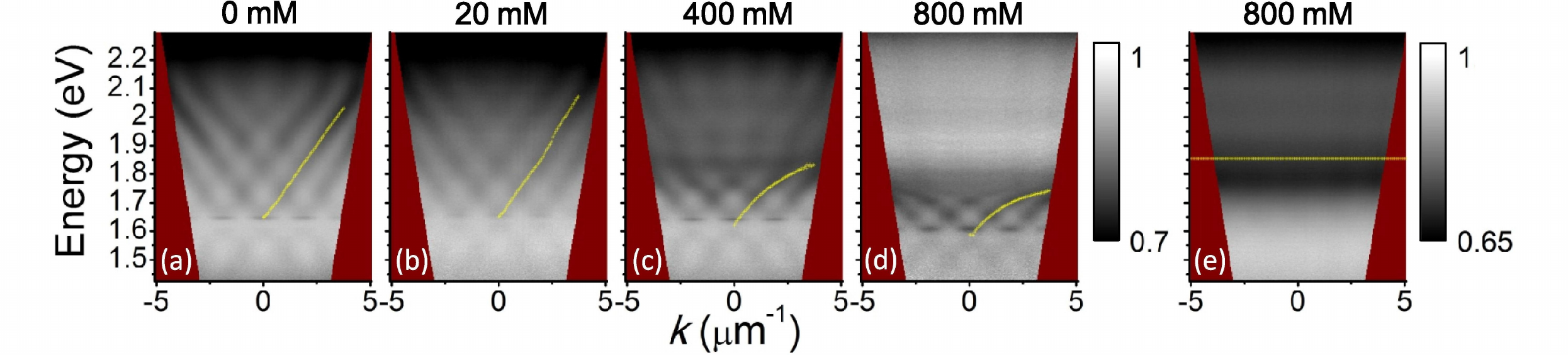}}
\caption{(a)--(d) The spatial coherence images for the array 2 with concentrations 0, 20, 400, and 800 mM, respectively. Here white areas correspond to transmission maximum. The yellow lines have the same meaning as in Fig. 2. (e) A sample having a random distribution of nanoparticles with 800 mM DiD concentration. Two transmission minima are seen at 1.85 eV (yellow line) and 2.25 eV, corresponding to DiD absorption and the single particle plasmon resonance, respectively.
}
\label{fig:spatialcoherence1}
\end{figure*}

In Figs.~\ref{fig:Dispersion}(q)--\ref{fig:Dispersion}(s), we plot the relative weights of the hybrid modes as functions of the in-plane wave vector $k$ for arrays 1--3, respectively, with molecular concentrations of 200, 400, and 800 mM. For arrays 1 and 2, the SLR-exciton hybrid is mostly SLR-like for $k\sim0$, and becomes increasingly excitonlike for higher $k$ values. The relative exciton contribution at $k\sim0$ increases with concentration due to stronger hybridization of the SLR with the exciton. Note, however, that for array 3 [Fig.~\ref{fig:Dispersion}(s)] the mode is excitonlike at $k\sim0$, and then gradually evolves to SLR-like mode at higher $k$. This is due to the SLR $\Gamma$-point energy being above the molecular excitation energy [compare, for example, Figs.~\ref{fig:Dispersion}(g) and~\ref{fig:Dispersion}(l)). These results demonstrate how the relative weights of the hybrid mode at a given energy and wave vector can be tailored by choice of geometry and molecular concentration.

To investigate coherence, angle resolved transmission spectra are recorded with a double slit placed on the image plane of the sample; see Fig.~\ref{fig:Setup}(b). This forms the crucial test for the presence of spatial coherence in the new modes: if the spatial coherence length of the mode is greater than the interslit distance, a distinctive fringe pattern would be expected in the Fourier plane of the imaging system. In Fig.~\ref{fig:spatialcoherence1}(a)--\ref{fig:spatialcoherence1}(d) are shown the wavelength-resolved spatial coherence images obtained from array 2 and with molecular concentrations ranging from 0 to 800 mM. Intriguingly, {\it bending of the interference pattern} is observed towards the strong coupling regime. In other words, one of the destructive interference fringes in spatial coherence images always overlaps with the extinction maxima of the dispersion (yellow symbols); see Figs.~\ref{fig:Dispersion}(g)-\ref{fig:Dispersion}(j). This allows to make an important connection with the original modes: If a spatially coherent light source (i.e., the sample) is radiating through a double slit, the interference fringes can be interpreted as replicas of the original dispersion (Fig.~\ref{fig:Dispersion}) created by the diffracted orders from the double slit. At high frequencies the interference pattern becomes complex due to the close spacing of the crossing points of the replicas (see also Supplemental Material \cite{Supplementarynote}). Thus, the fact that band bending with increasing concentration is seen both in the dispersions and the spatial coherence images provides a clear signature that the interference fringes are directly related to the modes of interest and are not due to any secondary reason. We have thus conclusively shown that the system modes have prominent spatial coherence throughout the crossover, also deep in the strong coupling regime. 

We want to point out the important role of the array periodicity, i.e\ the existence of the dispersive SLR modes, for the emergence of long-range coherence. Figure~\ref{fig:spatialcoherence1}(e) shows a spatial coherence image of a sample having a random interparticle spacing (for a SEM image, see Supplemental Material \cite{Supplementarynote}) while the molecular concentration, nanoparticle size, orientation and number are the same as in the sample in Fig.~\ref{fig:spatialcoherence1}(d). Evidently, no interference fringes are present in this case. Also, they are absent in DiD films without nanoparticles.

Notably, the fringes become less visible with increasing concentration at energies above 1.8 eV; see Figs.~\ref{fig:spatialcoherence1}(b)--\ref{fig:spatialcoherence1}(d). Higher molecular concentration induces stronger hybridization between the delocalized SLR and localized molecular excitons. At higher energies, these hybrid modes become increasingly excitonlike and localized as the energy gets closer to DiD dye absorption, reducing the spatial coherence length below the interslit distance. Note, however, that the fringe pattern persists below 1.8 eV energies, even with 800 mM concentration. We have thus demonstrated that the SLR-exciton hybrid modes display long-range coherence even when the mode is very excitonlike: from Fig.~\ref{fig:Dispersion}(r) the exciton weight can be deduced to be $80\%$ at high $k$-vector values.

\begin{figure}[ht!]
\centerline{\includegraphics[width=1\columnwidth]{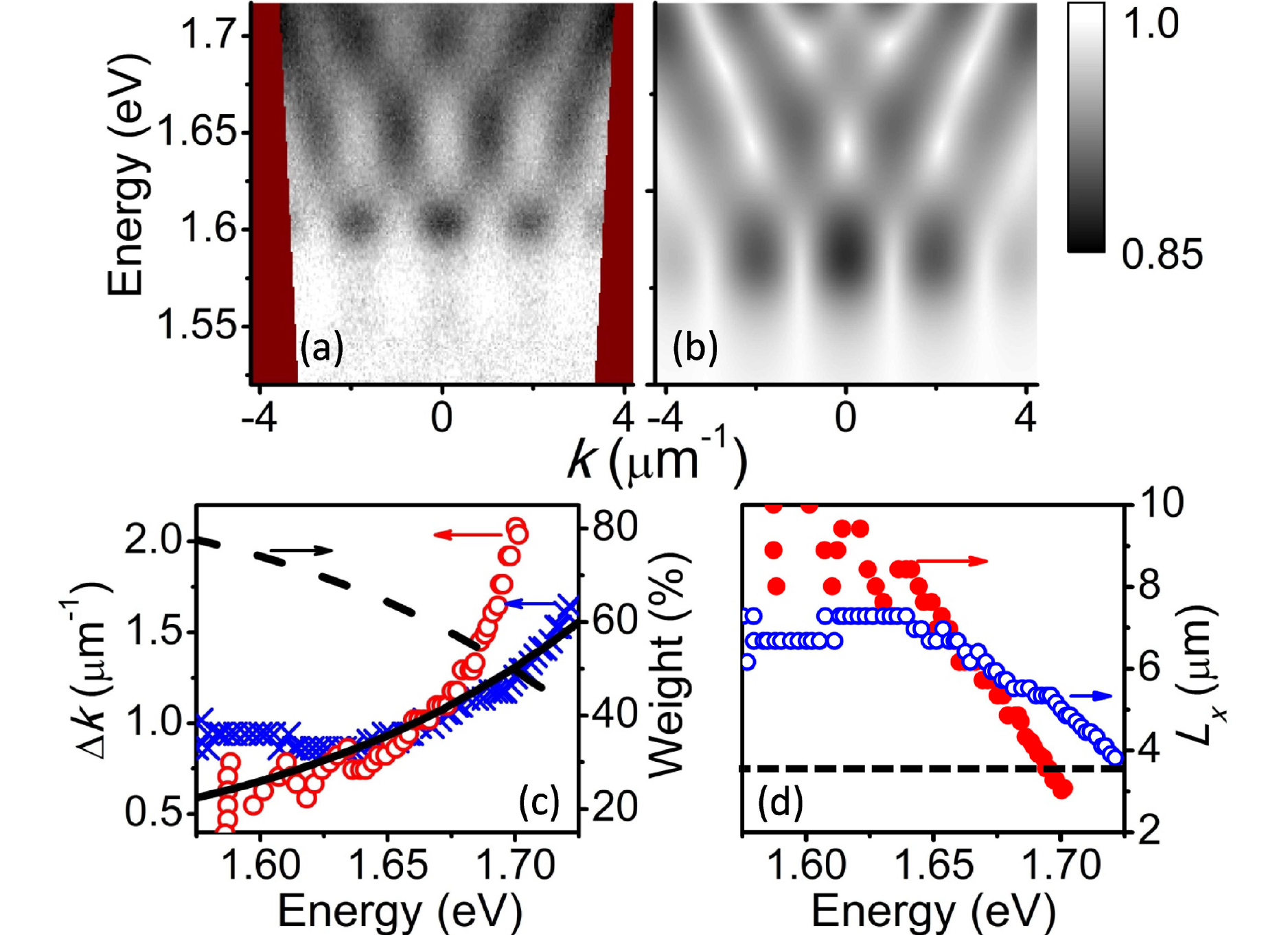}}
\caption{ (a) A close-up of the spatial coherence image (800 mM concentration). (b) The interference image obtained from the coupled oscillator model. (c) The $\Delta k$ obtained from the experiments (red empty circles) and from the coupled oscillator model (blue crosses). Dashed and solid lines correspond to the SLR and exciton weights of the mode, respectively. (d) The spatial coherence length obtained from the experiments (red circles) and from the coupled oscillator model (blue empty circles). The dashed line is the effective interslit distance at the sample plane.
}
\label{fig:spatialcoherence2}
\end{figure}

In the rest of this Letter, we consider the crucial question of whether there is a systematic, quantitative connection between the spatial coherence and the expected weight of the light component in a hybrid mode. First, we want to show that detailed structure of the interference fringes can be produced by assuming hybrid modes, with weights of the light and matter parts as obtained by fitting the experimental dispersion with the coupled oscillator model Eq.~\eqref{eq:matrix} (the obtained dispersion was then used to provide the energy and wave vector specific information of the mode radiating through the double slit, see Supplemental Material \cite{Supplementarynote}). In Fig.~\ref{fig:spatialcoherence2}(a) we show a close-up of the spatial coherence image of Fig.~\ref{fig:spatialcoherence1}(d) (800 mM concentration) and in Fig.~\ref{fig:spatialcoherence2}(b) we show the interference image obtained from calculations based on the coupled oscillators model. While the intensities in both Figs.~\ref{fig:spatialcoherence2}(a) and \ref{fig:spatialcoherence2}(b) are of comparable magnitude, at high energies, the experimental data have less transmission intensity. This can be due to additional absorption of the molecules that are not contributing to strong coupling \cite{Agranovich, Bellessa}. In general, however, the correspondence of the model with the most prominent features of the experimental data is excellent. This is the first step of systematically proving the connection between the hybrid structure and the coherence: the model with weights of matter and light parts in the hybrid as given by strong coupling theory indeed reproduces the interference pattern observed experimentally.

Second, we consider the important connection between the interference fringes, mode delocalization, the width of the mode $\Delta k$, and the relative weights of the strongly coupled modes. In Fig.~\ref{fig:spatialcoherence2}(c) we show the $\Delta k$ of the mode as a function of the energy obtained from the experiments (Fig.~\ref{fig:Dispersion}) and from the model. The $\Delta k$ was obtained as FWHM of constant-energy line cuts from the dispersions. Also shown are the relative SLR and exciton weights of the hybrid mode. In Fig.~\ref{fig:spatialcoherence2}(d) we show the spatial coherence lengths of the mode obtained as $L_x = 2\pi / \Delta k$ \cite{Mandel}. Because the momentum and position are Fourier related, a small $\Delta k$ at energies around 1.6 eV [see Fig.~\ref{fig:spatialcoherence2}(c)] suggests a delocalized mode and large spatial coherence length. The delocalization is also evident from the high SLR fraction (80 $\%$) of the mode. In the spatial coherence image, the delocalization manifests itself as a distinct interference pattern [Figs.~\ref{fig:spatialcoherence2}(a) and \ref{fig:spatialcoherence2}(b)]. As $\Delta k$ increases at energies $E > 1.65$ eV, however, the hybrid mode becomes more localized and more excitonlike, which gradually yields a less prominent interference pattern in accordance with the increasing weight of the matter component. At energies above 1.7 eV, the spatial coherence length decreases below the interslit distance [Fig.~\ref{fig:spatialcoherence2}(d)], and, consequently the interference pattern disappears; see Fig.~\ref{fig:spatialcoherence1}(d).

In both classical optics and quantum mechanics, modes are characterized not only by their energies, observable in dispersions, but also by the coherent modes or wave functions forming linear superpositions. Both aspects should be considered in identifying physical phenomena, cf.\ the observation of Bose-Einstein condensation by evidence in momentum distribution \cite{Anderson1995} and in interference patterns \cite{Andrews1997}. The strong coupling regime of various types of surface plasmon modes and emitters has been widely studied by observing dispersion relations. Splittings in the dispersions have been attributed to hybridization of plasmonic and excitonlike modes. Here we provide the first systematic study of the evolution of the spatial coherence in a plasmonic-molecule system when transiting from the weak to the strong coupling regime. The evolution of spatial coherence is shown to be directly connected to the hybrid mode structure. Significant spatial coherence lengths in the strongly coupled system are observed even when the mode is very excitonlike. Complementing the energy dispersions and dynamics observed earlier, our interference results provide conclusive evidence for the hybrid nature of the normal modes in strongly coupled surface plasmon---emitter systems. In general, our results demonstrate the potential of hybridization in creating nanosystems with designed properties, in this case long range coherence for modes that are largely matterlike.

\begin{acknowledgments}
We thank Dr.~Shaoyu Yin for useful discussions. This work was supported by the Academy of Finland through its Centres of Excellence Programme (Projects No. 251748, No. 263347, No. 135000, and No. 141039) and by the European Research Council (ERC-2013-AdG-340748-CODE). Part of the research was performed at the Micronova Nanofabrication Centre, supported by Aalto University.

\end{acknowledgments}


\newpage

\thispagestyle{empty}
\setlength{\hoffset}{-0.85in}
\setlength{\voffset}{-0.8in}

\includegraphics[page=1]{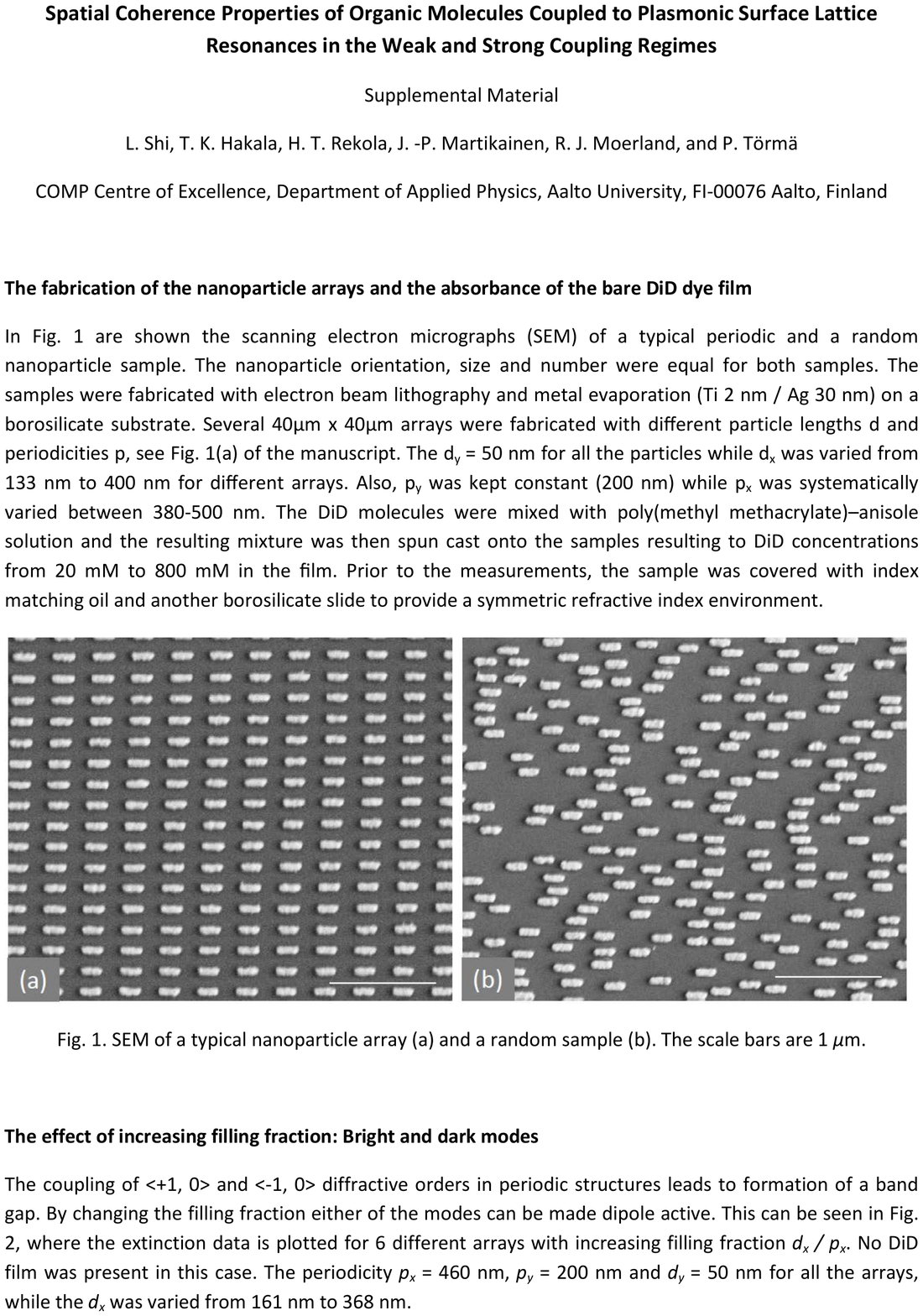}
 
\clearpage

\thispagestyle{empty}
\setlength{\hoffset}{-0.8in}
\setlength{\voffset}{-0.8in}
\includegraphics[page=2]{Supplemental_Material.pdf}

\clearpage

\thispagestyle{empty}
\setlength{\hoffset}{-0.8in}
\setlength{\voffset}{-0.8in}
\includegraphics[page=3]{Supplemental_Material.pdf}

\clearpage

\thispagestyle{empty}
\setlength{\hoffset}{-0.8in}
\setlength{\voffset}{-0.8in}
\includegraphics[page=4]{Supplemental_Material.pdf}
\clearpage

\thispagestyle{empty}
\setlength{\hoffset}{-0.8in}
\setlength{\voffset}{-0.8in}
\includegraphics[page=5]{Supplemental_Material.pdf}
\clearpage

\thispagestyle{empty}
\setlength{\hoffset}{-0.8in}
\setlength{\voffset}{-0.8in}
\includegraphics[page=6]{Supplemental_Material.pdf}
\clearpage

\thispagestyle{empty}
\setlength{\hoffset}{-0.8in}
\setlength{\voffset}{-0.8in}
\includegraphics[page=7]{Supplemental_Material.pdf}
\clearpage

\thispagestyle{empty}
\setlength{\hoffset}{-0.8in}
\setlength{\voffset}{-0.8in}
\includegraphics[page=8]{Supplemental_Material.pdf}
\clearpage

\thispagestyle{empty}
\setlength{\hoffset}{-0.8in}
\setlength{\voffset}{-0.8in}
\includegraphics[page=9]{Supplemental_Material.pdf}
\clearpage

\thispagestyle{empty}
\setlength{\hoffset}{-0.8in}
\setlength{\voffset}{-0.8in}
\includegraphics[page=10]{Supplemental_Material.pdf}
\clearpage

\thispagestyle{empty}
\setlength{\hoffset}{-0.8in}
\setlength{\voffset}{-0.8in}
\includegraphics[page=11]{Supplemental_Material.pdf}
\clearpage

\thispagestyle{empty}
\setlength{\hoffset}{-0.8in}
\setlength{\voffset}{-0.8in}
\includegraphics[page=12]{Supplemental_Material.pdf}
\clearpage

\thispagestyle{empty}
\setlength{\hoffset}{-0.8in}
\setlength{\voffset}{-0.8in}
\includegraphics[page=13]{Supplemental_Material.pdf}
\clearpage

\thispagestyle{empty}
\setlength{\hoffset}{-0.8in}
\setlength{\voffset}{-0.8in}
\includegraphics[page=14]{Supplemental_Material.pdf}
\clearpage

\thispagestyle{empty}
\setlength{\hoffset}{-0.8in}
\setlength{\voffset}{-0.8in}
\includegraphics[page=15]{Supplemental_Material.pdf}
\clearpage

\thispagestyle{empty}
\setlength{\hoffset}{-0.8in}
\setlength{\voffset}{-0.8in}
\includegraphics[page=16]{Supplemental_Material.pdf}
\clearpage

\thispagestyle{empty}
\setlength{\hoffset}{-0.8in}
\setlength{\voffset}{-0.8in}
\includegraphics[page=17]{Supplemental_Material.pdf}

\end{document}